\documentclass[prl,aps,groupedaddress,amsmath,twocolumn]{revtex4}

\usepackage[dvips]{graphicx}
\usepackage{dcolumn}
\usepackage{bm}
\usepackage[version=3]{mhchem} 

\begin{document}




\title{ The fate of the resonating valence bond in  graphene} 
\author{Mariapia Marchi}
\affiliation{SISSA, Via Bonomea 265, 34136 Trieste, Italy}\affiliation{Dipartimento di Fisica,
  Universit\`a di Trieste, strada Costiera 11, 34151 Trieste, Italy}
\author{Sam Azadi}
\affiliation{Institute of Physical Chemistry and                                                  
Center for Computational Sciences, Johannes Gutenberg University
Mainz, Staudinger Weg 9, 55128 Mainz, Germany}
\author{Sandro Sorella}
\email[]{sorella@sissa.it}
\affiliation{SISSA, Via Bonomea 265,
  34136 Trieste, Italy} \affiliation{Democritos Simulation Center
  CNR--IOM Istituto Officina dei Materiali, 34151 Trieste, Italy} 
\date{\today}

\begin{abstract}
We apply a variational wave function capable of  describing qualitatively 
and quantitatively the  so called ''resonating valence bond'' in realistic 
materials, by improving standard ab initio calculations by means
of quantum Monte Carlo methods.
In this framework we clearly identify the Kekul\'e and 
Dewar contributions to the chemical bond of the benzene 
molecule, and we establish  
the corresponding resonating valence bond energy of these well known structures ($\simeq 0.01$eV/atom).
We apply this method to unveil the nature of the chemical bond in undoped 
graphene 
and show that this picture remains only within a small ''resonance length'' 
of few atomic units.
\end{abstract}

\maketitle
Since the recent experimental isolation of 2D graphene
layers\cite{novoselov_science}, there has been a renovated
interest in the electronic properties of graphene. 
On the other hand the resonating valence bond  (RVB) theory was proposed 
several years ago by Linus Pauling\cite{pauling} and its successful application to 
aromatic compounds containing the benzene ring, has immediately raised the 
question whether this fascinating  theory 
remains meaningful in graphene, which can be viewed as a two dimensional realization of 
Carbon rings in a honeycomb lattice.   

Graphene is a subject of intense studies, also because its peculiar 
band structure implies a vanishing density of states at the Fermi energy 
with Dirac cones and non conventional semimetallic
behavior\cite{rise}.
We also mention that the photoemission properties, and a possible opening 
of a gap around the Dirac cones have not been fully understood neither
experimentally\cite{epitaxial_bandgap} nor
theoretically\cite{cinesi,insulator_tutti},
and recently it has been speculated that electron correlation may play a 
crucial role in this material\cite{baskaran}, and could lead not only to 
the explanation of this effect but also to a rather speculative    
$d+id$ (room)-high-temperature superconductivity upon doping. 
Generally speaking the role of electron correlation in graphene    
remains highly controversial\cite{castroneto}, and the  attention 
in the field has been renewed by a recent numerical simulation of the Hubbard 
model on the honeycomb lattice\cite{muramatsu}.
In that work, by using an  
unbiased numerical method, it was shown that 
the ground state of the model could be 
highly non trivial: an insulator, with neither  magnetic nor 
whatsoever broken symmetry, namely a RVB  spin liquid state.

In this Letter we clarify the role of RVB correlations in graphene and
other Carbon compounds by using a 
tool\cite{turborvb} for ab-initio calculations 
based on quantum Monte Carlo (MC) methods, capable 
of describing rather well  the electron correlation in 
several challenging molecules, up to the quantitative description of the 
weak binding in graphite\cite{spanu_graphite}.
With this technique we can visualize
the RVB character of the chemical bond, and describe 
realistically an RVB spin liquid state, with the same type of variational 
wave function  that has been shown to be rather accurate in model systems, 
especially in spin ones\cite{brian}. 
Since in realistic models that  allow charge fluctuations,  like e.g. the
Hubbard model, it is not possible to work with a complex  
wave function without breaking time reversal symmetry, we 
restrict our variational freedom  to real wave functions, which nevertheless 
allow a very wide class of spin-liquid states.

We shortly describe 
 the wave function used in this
Letter, as more details have been published elsewhere (see
e.g. \cite{casula_atoms,marchi} and refs. therein).
The RVB ansatz\cite{pwa}  $|\hbox{RVB}\rangle = J |\hbox{AGP}\rangle$ (JAGP) 
is made of a product of a Jastrow factor $J$, which
takes into account the short range strong Coulomb repulsion, and the 
so called antisymmetrized geminal power (AGP). A singlet valence bond between 
two electrons of opposite spin is determined by  
a geminal function $f$. At variance with the usual Slater Determinant (SD),  
where no correlation between opposite spin electrons is considered, 
in the AGP all the electrons are paired with the 
same geminal. The resulting wave function is then antisymmetrized.
We parametrize $f$ by using a given 
number $n^*$ of molecular orbitals (MOs) as
$f(\vec{r}^{\uparrow},\vec{r}^{\downarrow})=
\sum_{k}^{n^*}n_k \psi_k(r_{\uparrow})\psi_k(r_{\downarrow}),$
where $n_k$ are variational parameters.
The MOs $\psi_k$ are expanded in an atomic basis set and fully optimized 
by minimizing the variational MC (VMC) energy expectation
value of the full electron-ion  
Hamiltonian within the Born-Oppenheimer approximation\cite{marchi}.
In all the calculations of this work, we have replaced the $1s$ core electrons of 
the Carbon atom with appropriate pseudopotentials\cite{filippi_pseudo},
which also account for scalar relativistic effects.
When $n^*>N/2$, with $N$ denoting the number of electrons, the wave function 
has a larger variational freedom with respect to the best (lowest in energy) 
Jastrow SD wave-function (JSD)\cite{notelong}, 
and is able 
to improve the description of the electron correlation, especially
when the AGP
is used in combination with the Jastrow factor.  
The latter is particularly important for the description of a spin liquid 
state and is represented  by a weight factor  $ J( {\bf R})  = \exp [
\sum_{i < j} u(\vec r_i, \vec r_j)] $ over the $3N$--dimensional
configuration ${\bf R}$
of the electron positions $\vec r_i$. For the explicit form of $J$,
see e.g. Ref.~\onlinecite{marchi}.
Provided the two--electron function $u(\vec r_i, \vec r_j)$ decays slowly 
enough with the distance between the electrons $| \vec r_i - \vec r_j|$, 
it is possible 
to describe rather well a spin liquid insulator, 
even when, in absence of $J({\bf R})$, the AGP pairing function describes a semimetal (for $n^*=N/2$) or a superconductor (for $n^*>n$)\cite{capello}. 
As discussed in Ref.~\onlinecite{marchi},
an appropriate choice of $n^*$ is crucial to improve the accuracy in
the description of the chemical bond with the JAGP ansatz:
$n^*$ is the minimum number of MO's  that can be used   
for describing a product of independent 
Hartree-Fock (HF) wave functions for isolated atoms.
Within this choice of $n^*$, both the two--electron functions $f$ and $u$ are expanded in a basis of localized gaussian atomic orbitals, with a method that in principle converges
to the complete basis set limit (CBS), 
yielding the lowest possible energy state compatible 
with the given ansatz\cite{azadi}. 

We test our variational ansatz on small
Carbon compounds. We consider  
the atomization energy of the Carbon dimer and of benzene, 
computed  as the difference between the JAGP energy for the
entire molecule and the JSD energy of the isolated atoms\cite{marchi}.
To compare our results with the experimental binding energies we
also include inner shell correlations and relativistic effects and we
subtract the zero-point energy. 
In Table~\ref{c2_c6h6_binding} we show our VMC and lattice-regularized
diffusion MC (LRDMC)\cite{lrdmc} results.
The simulations for  the Carbon dimer were performed in
Ref.~\onlinecite{marchi} (n$^*$=7). We evaluate inner shell
correlations by comparing the all electron \ce{C2} energy found in
Ref.~\onlinecite{umrigar_diatomic} with the energy found in
Ref.~\onlinecite{sorella_umrigar_prl}, where the same
pseudopotential\cite{filippi_pseudo}  as for Ref.~\onlinecite{marchi}
was used. We take spin orbit effects from
Ref.~\onlinecite{bytautas}. 
For benzene (n$^*$=24) 
we use a large basis set close to the  CBS--limit within   
0.01eV/atom in all the cases studied.
We take inner shell correlation and spin orbit effects from
Ref.~\onlinecite{parthiban}.
\begin{table} 
\begin{center}
\begin{tabular}{|l|l|l|l|l|l|}
\hline
Molecule& (V)JSD & (V)JAGP& (LR)JSD & (LR)JAGP&Exp.\\
\hline 
\ce{C2} & 5.54(2) & 6.33(2) & 5.76(2) &6.30(2) &6.30(2)$^a$ \\ 
\ce{C6H6} & 56.98(1) & 57.11(3) & 57.11(1)&57.14(1)&56.62(3)$^b$   \\ 
\hline
\multicolumn{6}{l}{$^a$ Ref.~\onlinecite{bytautas}, $^b$ Ref.~\onlinecite{parthiban}}\\
\end{tabular}
\caption{VMC (V) and LRDMC (LR) atomization energy (in eV) of \ce{C2}
  (AGP primitive basis:   $5s5p$) and \ce{C6H6} (\ce{C} primitive basis:
  $24s22p10d6f$; \ce{H}: $3s2p$). 
   }
\label{c2_c6h6_binding}
\end{center}
\end{table}
One of the main results of our calculation is 
represented by the sizeable energy gain that is obtained by using a 
large number of 
MOs in the AGP part of the wave function.
This energy gain is 
particularly important to get a quantitative description of the 
\ce{C2} chemical bond, whereas it is possible that the 
slight overestimation of the  atomization energy  for benzene 
does not depend on the accuracy of our total energy estimates, but comes out 
from the previously 
described corrections taken from other methods/experiments. This is 
plausible considering that 1) the result provided by LRMDC, which is known to improve the energy estimate\cite{lrdmc}, is 
in  agreement with our best variational ansatz, and that 2) 
we have {\em never} overestimated the well depth by more than 
$0.01$eV/atom in all the cases studied in the
previous work\cite{marchi}. 
Anyway, our variational ansatz appears to be adequate  
and encourages us  
to quantify the  
RVB energy, which, in the present
formulation, can be defined as  
the energy difference between the best variational energy found 
with $n=N/2$  MOs and the one with $n=n^*>N/2$, both obtained in presence of 
$J$. 
In Table~\ref{c6h6_pi_tab}, we report the contribution of the
$\pi$-band orbitals to the RVB energy of benzene and graphene. The $\pi$
orbitals yield approximately the 80$\%$ of the pairing 
and represent in general the most important contribution, as expected.
\begin{table} 
\begin{center}
\begin{tabular}{|l|l|l|l|l|}
\hline
%
Molecule & \ce{C6H6} & \ce{8C} &\ce{16C} &\ce{48C}\\
\hline 
All & 0.118(2) &  0.159(7) & 0.207(4) & 0.18(1) \\ 
\hline
$\pi$ & 0.101(2)& 0.116(5)&0.147(8)& 0.15(1) \\
\hline
\end{tabular}
\caption{VMC contribution (in eV) of all (All) the occupied bands and
  of the $\pi$ band to the binding energy of \ce{C6H6} and 
  graphene layers of \ce{8C}, \ce{16C}, and \ce{48C} atoms 
 (AGP primitive basis: $11s9p7d$).}
\label{c6h6_pi_tab}
\end{center}
\end{table}

\begin{figure}
\centering
\includegraphics[width=1.0\columnwidth]{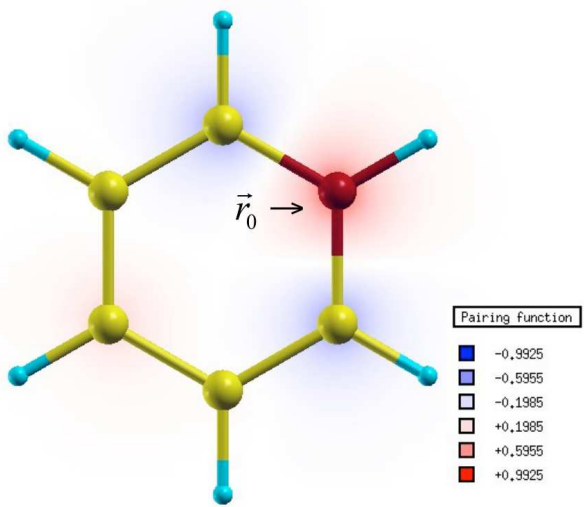}
\caption{
\label{kekule}
(Colors online) Two dimensional plot of the AGP pairing function
restricted to the molecular orbitals above the HOMO. 
The arrow indicates the reference position $r_\uparrow$ fixed on an
atom, colored in red for the sake of clarity. 
} 
\end{figure}

To get a deeper insight into our variational calculation 
with the JAGP wave function, 
we introduce also  a ``valence-projected pairing function'' (VPPF),
defined as
$f_{VPPF}(r_{\uparrow},r_{\downarrow})=\sum_{k>N/2}^{}n_k \psi_k(r_{\uparrow})
\psi_k(r_{\downarrow}).$
In the HF case of a single SD,
$f_{VPPF}(r_{\uparrow},r_{\downarrow})=0$. Hence, when singlet valence 
bond  pairing occurs  and $n_k$ is non zero even for $k> N/2$, we can 
visualize and characterize, in real space, the genuine RVB contribution 
to the chemical bond. 
We can also plot the VPPF restricted to
the $\pi$-band as a function of $r_\downarrow$ as done for benzene in
Fig.~\ref{kekule}. 
Kekul\'e nearest neighbor and Dewar further neighbor 
correlations are manifest. Fig.~\ref{kekule} proves 
the JAGP wave function to be a powerful tool for the description of  
the fundamental features of the RVB  chemical bond.

\begin{figure}
\centering
\includegraphics[width=1.\columnwidth]{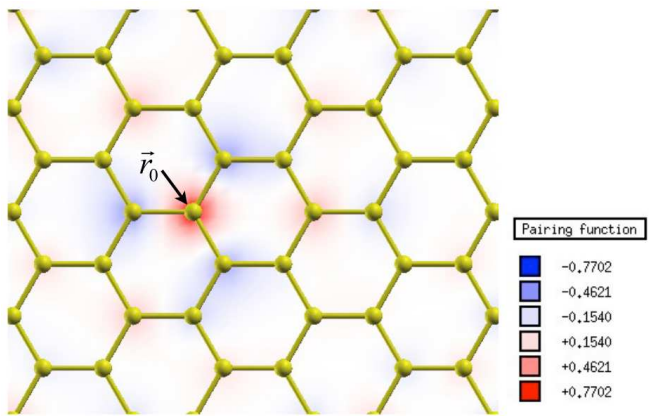}
\caption{
\label{c48}
(Colors online) Two dimensional plot of the AGP pairing function for a graphene layer
of 48 C atoms restricted to the molecular 
orbitals above the HOMO  (VPPF).  The 
arrow indicates the reference atom. } 
\end{figure}

\begin{figure}
\centering
\includegraphics[width=0.9\columnwidth]{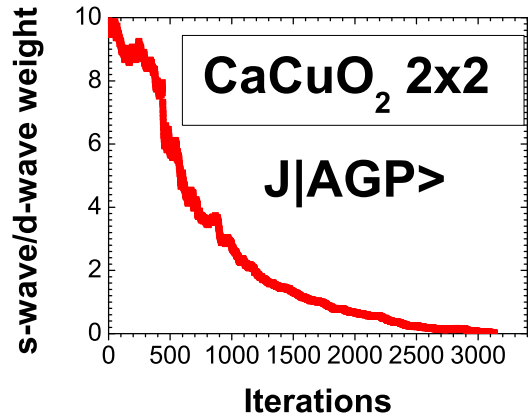}
\caption{(Colors online) Ratio between the  $s$-wave and the  $d$-wave
weight in the JAGP wave function for the \ce{CaCuO2} parent compound of 
the high-temperature cuprate superconductor (2x2 supercell).} 
\label{sd}
\end{figure}

We  now discuss the 
case of undoped graphene.
We consider rectangular supercells $L_x \times L_y$, with  
$L_x = 3 n  a$ and $L_y=m \sqrt{3} a$ where $a=1.42$ {\AA}~is the nearest 
neighbor Carbon distance and $n,m$ are integers. 
We use an increasing
number $4 n m$ of \ce{C} atoms (8, 16, 24, and 48, with $n,m$ such that 
$L_x/L_y \simeq 1$). These supercells do 
not satisfy the $\pi/3$ rotation symmetry of the 
infinite lattice\cite{bernu}.  
This helps the system to break rotational symmetries, 
such as $d_{xy}$ or $d_{x^2-y^2}$ for a real pairing function, 
that are  energetically favored  when  the  expected 
$d+id$ pairing symmetry\cite{doniach}  characterizes 
the ground state wave function.
For each
system size, we optimize the JAGP wave function, find the VMC
energy and the VMC RVB energy (by means of correlated--sampling
simulations). As reported in Table~\ref{c6h6_pi_tab}, we 
have also checked the contribution of the
$\pi$--band orbitals to the RVB energy gain. 
In Fig.~\ref{c48} we show the VPPF
(restricted to the $\pi$--band) for the 
largest  supercell considered here. Despite  
the small number of atoms,  we 
already see an almost perfect rotational symmetry of the VPPF, that is  
not compatible with $d$--wave pairing. 
To prove that 
our method is capable of tackling with pairing functions with   
$d-$wave symmetry  
we apply our scheme to
the \ce{CaCuO2} parent compound of cuprate high-temperature superconductors. 
As shown in Fig.~\ref{sd}, in less than 3000 iterations
we melt the $s$--wave pairing and are able to detect the correct 
$d$--wave symmetry of the pairing function. 
We can conclude, therefore, that the 
RVB chemical bond in graphene is characterized 
by a pairing function with a clear $s$--wave symmetry.

\begin{figure}
\centering
\includegraphics[width=1.0\columnwidth]{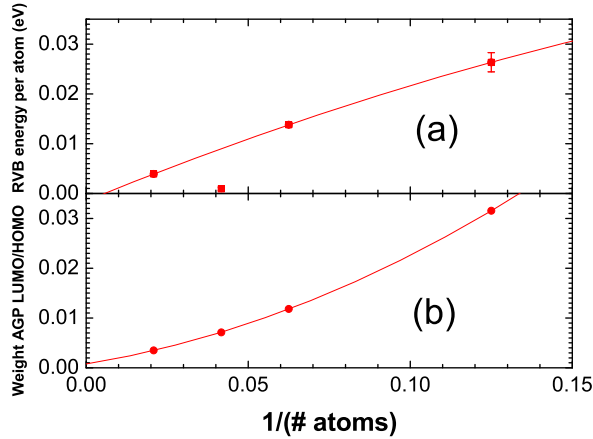}
\caption{
\label{theendofrvb}
(Colors online) (a): 
RVB energy per atom for the 
graphene layer, as a function of
the number of atoms in the supercell ($\Gamma$ point). (b):  
Ratio of the LUMO/HOMO  weight in the AGP, a measure of the RVB character 
of the bond.  Lines are guides to the eye.
} 
\end{figure}

Finally, in order to understand the thermodynamic properties 
of graphene, we consider a finite size scaling of our results.
In Fig.~\ref{theendofrvb} we show the energy gain due to the $s$--wave RVB
(upper panel) and the ratio $R= n_{N/2+1}/ n_{N/2} $ 
of the LUMO/HOMO weights $n_k$ as a function of the inverse 
 number of \ce{C} atoms in the supercell. 
Before discussing this result, we recall what happens 
to the above--mentioned quantities in the absence of correlation, i.e. when there is 
no Jastrow factor in our variational ansatz. In such a case,
if the ratio $R$ converges to a finite quantity in the thermodynamic limit, 
the AGP wave function describes  an  $s$--wave 
superconductor with true off diagonal long 
range order. Besides, the  thermodynamic--limit RVB energy per 
atom remains finite and  represents just the  
condensation energy of the $s$--wave superconductor.
In the presence of $J$ 
instead, a different scenario is
possible. Indeed, 
a ratio $R>0$ in the thermodynamic limit and a finite RVB energy/atom 
denote a spin liquid state
with a spin and a charge gap in its spectrum. 
This possibility is 
compatible with the recent Hubbard model results\cite{muramatsu}, 
and may explain also the existence of a small gap in the photoemission 
experiments, genuine and determined  only by the RVB character of 
the ground state. 
Due to the computational cost increase for larger supercells,
it is difficult 
to obtain  an accurate thermodynamic limit  with our VMC method. 
However, clear trends are 
evident from 
Fig.~\ref{theendofrvb}. In the upper panel, we see that the energy
gain of the RVB wave function systematically decreases
as the system size increases, apart for the negligible value found for the
\ce{24C} supercell. 
The anomaly of the \ce{24C} cluster can be easily
explained as a shell--effect. Indeed, this cluster should be closer to the
thermodynamic limit, since it contains the
so-called $K$ point, the gapless Dirac point in graphene.
This shell effect does not affect the eigenvalues of the
pairing function, which instead decrease monotonically as the system
size increases and reach a 
very small value in the thermodynamic
limit (lower panel). 
If we extrapolate the upper--panel results, omitting the \ce{24C} cluster, 
also the RVB energy per \ce{C} atom becomes extremely 
small in the thermodynamic limit (smaller
than the accuracy of the present data).
Both panels thus suggest that the semimetal character of graphene 
should be stable in the thermodynamic limit. 
A small gap could appear in the excitation spectrum 
only if its value was extremely small $\simeq 0.01$eV.
We have estimated this value by matching our results for the $n_k$ with the 
ones obtained with an $s$--wave BCS hamiltonian with nearest and next-nearest 
neighbor coupling, describing a $Z_2$ gapped spin liquid\cite{cinesi} 
 when  correlation 
is included by means of an appropriate Jastrow. 

In conclusion we have systematically studied Carbon--based compounds from 
the simplest \ce{C2} molecule to graphene layers. We have shown that the RVB 
character of the chemical bond can be depicted in terms of a very powerful 
and accurate wave function that not only improves the 
description of the chemical bond but it is 
 also capable to show qualitatively 
new effects induced by the electron correlation.
We have found clear numerical evidence 
 that singlet $s$--wave pairing in graphene be quite 
robust and sizeable up to a small length scale of few atomic units.
This feature might remain in the thermodynamic limit 
leading to a very small 
gap in the photoemission spectrum or to $s$--wave superconductivity upon 
doping, effects that can be in principle verified experimentally.

\acknowledgments
This work is supported by CINECA (PEC grant 2008) and MIUR (COFIN 2007).
We acknowledge useful discussions with A. Morgante, F. Becca and  
F. Mauri.





\begin{thebibliography}{99}
\bibitem{notelong} In particular our parametrization allows us to represent 
also slowly decaying pairing functions, as for instance the one obtained 
within the Hartree Fock theory, namely $f(|r -r^\prime|) \propto | r -r^\prime|^{-2}$.
\end{thebibliography}
\end{document}